\documentclass[proof]{WileyASNA-v1}
\articletype{Article Type}%
\received{5 December 2019}
\revised{}
\accepted{}
\begin{document}
\title{Characterization of Variability in Blazar Light curves}

\author[1,2]{K. K. Singh*}

\author[1]{P. J. Meintjes}

\authormark{K. K. Singh \& P. J. Meintjes}

\address[1]{\orgdiv{Physics Department}, \orgname{University of the Free State}, \orgaddress{\state{Bloemfontein}, \country{South Africa}}}

\address[2]{\orgdiv{Astrophysical Sciences Division}, \orgname{ Bhabha Atomic Research Centre}, \orgaddress{\state{Mumbai}, \country{India}}}

\corres{*K. K. Singh, \email{kksastro@barc.gov.in}}

\presentaddress{Physics Department, University of the Free State, Bloemfontein- 9300, South Africa}

\abstract{Blazars represent dominant population of the extragalactic $\gamma$-ray sources in the Universe. 
These sources exhibit some characteristic properties like strong and non-thermal continuum emission over 
the entire electromagnetic spectrum from radio to TeV $\gamma$-rays with rapid variability on all timescales. 
The emission at radio and optical wavelengths is highly polarized with significant variation. The fastest variability 
in the blazar emission is observed during the flaring activity which is an important observational property of blazars. 
In this paper, we describe various methods to characterize the temporal variability in the multi-wavelength light curves 
of blazars. We also provide a detailed description of the set of statistical parameters which are used to quantify the 
level of variability present in the time-series. Implications of the informations derived from the variability study 
to probe the physics of blazars using multi-wavelength observations are also discussed.}


\maketitle


\section{Introduction}\label{sec1}
Blazars are classified as radio-loud active galactic nuclei (AGN) with relativistic plasma jets 
oriented at small angles to the line of sight of the observer at the Earth. The oppositely directed 
plasma jets in blazars originate from the central region of the elliptical host galaxy and linearly 
extend up to Mpc scale in the intergalactic medium \citep{Begelman1984,Urry1995}. The large bolometric 
luminosity (up to 10$^{49}$ erg~s$^{-1}$) measured from blazars suggests high accretion rate onto a 
supermassive black hole (SMBH; 10$^6$-10$^{9}$ M$_\odot$) at the centre of the host galaxy. 
It is commonly accepted that relativistic jets are powered by the rotational energy of a spinning 
SMBH and a magnetized accretion disk \citep{Blandford1977,Blandford1982}. The relativistic jets transport a 
large amount of energy and momentum over Mpc scales and dissipate a small fraction of energy during 
their evolution. The interaction of energetic particles in the jet plasma outflow with the ambient matter, 
radiation and magnetic field produces continuum emission over the entire electromagnetic spectrum. 
However, the matter content of the jet and physical processes responsible for launching, collimating and 
accelerating the jets have not been clearly identified until today. The multi-wavelength radiation from 
radio to TeV $\gamma$-rays measured from blazars is described as the non-thermal emission of relativistic 
plasma outflow along the jet axis. It is believed that the emissions from blazar jet are highly beamed due 
to its peculiar orientation close to the line of sight of the observer. Therefore, the broadband continuum 
radiation from the jet is strongly amplified by the relativistic Doppler boosting due to the high values of 
the apparent Lorentz factor of the emission region. The Doppler factor ($\delta$) of the emission region 
is given by 
\begin{equation}\label{Doppler-factor}
	\delta~=~\frac{1}{\Gamma_b(1-\beta_b~\rm cos~\theta)}
\end{equation}
where $\Gamma_b$ is the bulk Lorentz factor, $\beta_b$ is the speed of emission region in units of speed of light ($c$) 
and $\theta$ is the angle between jet axis and line of sight of the observer. In the small viewing angle approximation 
($\theta~\le~10^\circ$) for blazars, $\delta \approx \Gamma_b$, and non-thermal emission from blazars is Doppler boosted 
by a factor of $\delta^4$ due to relativistic beaming. The isotropic luminosity ($L_{iso}$) inferred from a blazar at 
redshift ($z$) is related to the intrinsic luminosity ($L_{int}$) emitted from the sources as 
\begin{equation}
	L_{iso}~=~\frac{\delta^4}{(1+z)^2}~L_{int } 
\end{equation}	
The low energy emission at radio, IR-optical-UV and soft X-ray energies is produced by the synchrotron radiation of 
relativistic electron-positron pairs in the magnetic field. The measurement of high degree of linear polarization at 
radio/optical wavelengths indicates the presence of ordered magnetic field in the jet emission region \citep{Singh2019a}. 
Dedicated radio observations of blazars over wide frequency range indicate the presence of static and dynamic shocks in 
their jets \citep{Aller1985,Marscher2010}. The presence of shocks in the localized regions in the jet is assumed to be 
responsible for the acceleration of particles to relativistic energies via \emph{Fermi} acceleration 
process \citep{Rieger2007}. The high energy emission in the hard X-ray and $\gamma$-ray bands is explained in the frame-work 
of two alternative scenarios based on leptonic and hadronic processes \citep{Bottcher2013}. In the leptonic scenario, 
the inverse Compton scattering of the synchrotron photons produced in the jet or thermal photons from the radiation fields 
external to the jet by the relativistic electrons/positrons is invoked to describe the observed high energy emission from the 
blazars \citep{Tavecchio1998,Bottcher2007,Dermer2009}. The hadronic models assume that relativistic protons emit high energy 
radiation directly through synchrotron process or via photo-hadronic interactions resulting in meson decay and particle 
cascades \citep{Mannheim1993,Aharonian2002,Dermer2012}. The long term correlated variability between low and high energy 
emissions observed from blazars generally supports the leptonic models \citep{Zhang2018} whereas detection of astrophysical 
TeV neutrino events favors hadronic scenarios \citep{Xavier2018}. However, the exact physical process for the high energy 
emission from blazars has not yet been identified and remains an open problem in blazar research.
\par
Statistical study of the variability in the multi-wavelength light curves of blazars provides a unique tool to distinguish 
the physical mechanisms involved in the high energy emission from these sources. The X-ray and $\gamma$-ray emissions from 
majority of the blazars are observed to be more variable than low energy emissions and dominate in their broadband 
spectrum energy distributions. The nature of variability observed in blazar emissions is stochastic and occasionally leads 
to the production of strong flares \citep{Singh2012,Liodakis2018}. During the flaring episodes, a dramatic increase in 
the flux level is detected from these sources. Flaring activity is observed either simultaneously in different energy bands 
or with a time-lag. Orphan flares in a given energy band without any counterpart in other wavebands are also frequently 
detected. Such activities in blazars are very important to identify the dominant emission mechanisms because 
X-ray and $\gamma$-ray flares are generally accompanied by a significant change in the properties of the optical 
polarization \citep{Abdo2010a,Singh2019a}. Therefore, proper characterization and quantification of the variability 
observed in the blazar light curves during low and high activity states are essential to probe the physical processes 
operating in the jet emission region. In this paper, we discuss the characterization and quantification of the 
multi-wavelength variability in the blazar time-series using various techniques and statistical parameters. The structure of 
the paper is the following. The variability observed in blazars is briefly reviewed in Section 2. In Section 3, we describe 
the methods for characterizing the temporal variability. The statistical parameters used for quantifying the level of variability 
are presented in Section 4. Finally, we discuss and conclude the importance of variability study in Section 5.


\section{Blazar Variability}\label{sec2}
Variability observed in the multi-wavelength emissions from blazars at different timescales is considered as one 
of the most important defining characteristic of these extragalactic sources. During the violent flaring activity of 
the blazars, the flux levels are observed to increase by a large factor over different timescales ranging from months to 
few minutes. The fastest variability at minute timescale is generally observed in the X-ray and $\gamma$-ray light curves 
of the blazars \citep{Cui2004,Aharonian2007,Albert2007,Ackermann2016,Singh2018,Singh2019b}. The characteristic variability 
timescale in the co-moving source frame ($t_{source}$) is shortened due to relativistic Doppler effect and can be estimated 
from the observed variability timescale ($t_{obs}$) using the relation, 
\begin{equation}
	t_{source}~=~\frac{\delta}{(1+z)}~t_{obs}
\end{equation}	
implying relatively faster variability in the observed emissions. Therefore, the blazar variability is generally divided into 
the following three classes according to the observed variability timescale:
\begin{itemize}
	\item{\bf Long Term Variability (LTV):}   $t_{obs}$ ranges from months to several years or decades. 

         \item{\bf Short Term Variability (STV):} $t_{obs}$ between few days to weeks or months.

	\item{\bf Intra Day/night or Micro- Variability (IDV):} $t_{obs}$ from minutes to hours or less than one day. 
\end{itemize}
The blazars are observed to randomly exhibit any of the above variability during flaring episodes at different epochs 
but there is no clear explanation for the stochastic nature of the blazar variability. The micro-variability at 
minute timescale (generally observed in X-ray and $\gamma$-ray light curves) suggests that the origin of flux variations 
lies in the jet where particles are accelerated to relativistic energies. The probable intrinsic mechanisms in the jet  
are  magnetic irregularities, relativistic mini-jets or changes in the viewing angle of the discrete emission regions.  
Whereas, the shape of flux distribution during flaring episodes suggests that the variability originates due to 
multiplicative processes associated with the accretion disk. Extrinsic effects like rapid swing in the jet viewing angle 
or changing Doppler factor due to bending in the jet and gravitational lensing are also invoked to explain the blazar variability.
Following models based on above physical processes have been proposed to explain the observed variability in the multi-wavelength 
light curves of blazars under leptonic frame-work. 

\subsection{Shock in Jet Model}
Internal shocks moving along the jet are proposed to explain the observed variability of blazars at different timescales 
\citep{Marscher1985,Spada2001}. This model is mainly based on the assumption that the energy injected by the central 
engine of blazar is channeled intermittently to accelerate the plasma shells in the jet. The collision of accelerating plasma 
shells with different velocity, mass and energy leads to the development of energetic shocks in the jet. The ordered bulk kinetic 
energy of these internal shocks propagating along the jet is partially converted into the magnetic field and radiative energy 
output of the particles. A time dependent model for the shell-shell collision causing the production of internal shocks 
can explain the observed broadband correlated variability in blazars at different timescales. This model requires a relatively low 
radiative efficiency for the dissipation of bulk kinetic energy which is consistent with the kinematics of blazars. Inhomogeneous 
collision of very thin shells consisting of energetic electrons and positrons and propagating along the jet axis with relativistic 
speed also causes flaring activities in the jet \citep{Blazejo2000}.

\subsection{Fluctuating Accretion Disk Model}
Fluctuations in the accretion disk due to variations in the accretion rate near the inner radius causes variability in 
flux \citep{Lyubarskii1997}. The changes in the accretion rate are attributed to the independent spatial fluctuations 
in the viscosity at radii larger than the inner radius of the disk. The energy release near the inner radius of the disk 
can lead to separate flaring activity due to magnetic reconnection. This model is generally applicable for the X-ray variability 
from the blazars with timescales longer than the light crossing time across the Schwarzschild radius of the SMBH in the central region 
of host galaxy. The log-normal distribution of flux points in the light curve during flares suggests multiplicative processes associated 
with the fluctuating accretion disk models \citep{Arevalo2006}. The stochastic nature of the variability can be attributed to modulated 
emission by the disk of SMBH.

\subsection{Spine-Layer Model}
A spine-layer configuration for blazar jet is proposed to explain the variability observed in different energy 
bands \citep{Ghisellini2005,Tavecchio2008}. In a structured jet, it is assumed that a fast jet spine (core) is cospatially 
surrounded by a slow layer (sheet). The slow layer can be produced through the interaction of the walls of the jet with the ambient 
medium or due to the acceleration of a discontinuous but intermittent jet. This scenario is similar to the internal shock or 
shock in jet model discussed above. A strong radiative interplay and feedback exist between the layer and spine. The 
high energy emission takes place in the layer whereas the spine is responsible for the low energy radiation observed from 
the blazars. At small viewing angles, the Doppler boosted emission from the spine dominates whereas the layer with broader beaming 
contributes to the over all emission from the blazar.

\subsection{Needle-in-a-Jet Model}
This model is analogous to the spine-layer model to avoid very large values of the bulk Lorentz factor ($\Gamma_b~\sim$ 50) and narrow 
jet viewing angles ($\theta~\sim~1^\circ$) for the rapid variability in blazars \citep{Ghisellini2008}. In this model, very small active 
regions move inside the blazar jet, with a speed faster than the rest of plasma. The compact active region is referred to as a needle in 
the larger jet. The needle traveling in a dense radiation field of jet emission is filled with the tangled magnetic field similar to 
the jet. The flaring episodes with rapid variability are produced by the fast moving needles of plasma, oriented in different directions 
inside a wider cone and occasionally aligned with the line of sight.

\subsection{Minijets-in-a-Jet Model}
In this model, the rapidly varying multi-wavelength emission originates from a very compact and relativistically moving emission 
region within a Poynting flux dominated jet of $\Gamma_b~\sim$ 10 \citep{Giannios2009}. The magnetic reconnection events in the 
highly magnetized regions dissipate a small fraction of the total jet luminosity. The plasma outflows from the magnetic 
reconnection regions of the jet power the $\gamma$-ray flares through the inverse Compton scattering of the synchrotron photons. 
The emission region in the larger jet is referred to as a minijet. The minijets-in-a-jet statistical models are used to 
interpret the fast variability and statistical properties of the $\gamma$-ray light curves \citep{Biteau2012}. This model 
suggests that the fast variability in high energy emission originates from the jet through additive processes unlike the 
multiplicative processes associated with the disk. The observed flux from a blazar is sum of the contributions from several 
independent randomly distributed minijets in an optically thin medium. The addition of individual components with finite moment 
leads to a log-normal distribution of the observed flux points following the central limit theorem.

\subsection{Relativistic Turbulence Model}
This model is inspired from the minijets-in-a-jet model and invokes relativistic turbulent processes to explain the flaring 
activity of blazars with short variability timescales \citep{Narayan2012}. In this model, various minijets are assumed to be 
relativistically moving in random direction and the variability timescale is estimated from the size of these emission 
regions in their comoving frames. The random motion of emission regions is a result of the relativistic turbulence generated 
due to magnetohydrodynamical plasma instability in the blazar jet. The radiation is observed from these dynamic turbulent 
emission zones for a short duration (given by the longitudinal size of the nearly spherical blobs) when their direction of 
motion points in line of sight of the observer. 

\subsection{Jet-Star Interaction Model}
This model is proposed for the variability at minute-scale during TeV flares on top of the $\gamma$-ray emission 
varying at timescales of days \citep{Barkov2012}. According to this model, ultra-short TeV flares are produced by 
the plasma condensation due to interaction between red giant stars and base of the jet close to central SMBH. 
The emission model is based on the time dependent evolution of the envelope lost by the red giant star in the jet 
and its non-thermal emission through leptonic and hadronic processes. The powerful blazar jets with high ram pressure 
can drag and disrupt the star atmosphere or envelope into ensemble of compact magnetized plasma condensates generally 
known as blobs. The non-thermal emission from these relativistically moving blobs with bulk Lorentz factor up to 100 can 
exhibit variability at short timescales.   

\subsection{Helical Jet Model}
This is a general relativity based model of the variability in the emissions from the blazar jet \citep{Mohan2015}.
In this model, a plasma flow is assumed to transit from the accretion disk onto the jet through magnetic field surface 
near the central SMBH. In the radiation pressure driven flow, the density inhomogeneties or blobs can be present at 
various scales in the mass-loaded jet. These orbiting blobs move relativistically along the bulk plasma in the region with 
a dominant magnetic field. In this region, the plasma is constrained by flux freezing to follow the motion along the magnetic 
surface encompassing the jet. The radiation originating from these orbiting blobs in helical motion along the jet is beamed in 
the line of sight of the observer and it causes rapid variations in the observed flux. 

\subsection{Multi-zone Emission Model} 
Multi-zone models are based on the internal structure of the emission region with more than one dissipation zones in the 
blazar jet. Multiple radiation zones are created by the collisions of stationary and moving shocks in the jet. 
Multi-zone emission models are used to explain the short term multi-wavelength variability in blazar light curves. Single 
zone models with leptonic or hadronic processes are useful for long term variability since the injection time into the emission 
region with finite size and light crossing effects predict a larger minimum timescale for flaring episodes \citep{Eichmann2012}.
However, time-dependent single zone models with several injections of the radiating particles into the emission region can 
explain the multi-wavelength blazar flares of short duration \citep{Roken2018}. These models are different from the one-zone time dependent 
models with single injection for the blazar variability at different timescales \citep{Mastichiadis2013,Singh2017}.
  
\subsection{Perturbation in Particle Acceleration Model}
This model is based on the small temporal fluctuations in the acceleration of particles responsible for non-thermal emission from 
the blazars \citep{Sinha2018}. According to this model, electrons are accelerated at a shock front and subsequently diffuse in 
the cooling region. A small perturbation in the acceleration timescale causes variations in the number density of accelerated electrons 
in the cooling or emission region. Variation in the escape time of electrons in the acceleration region introduces non-linearity in 
the electron distribution in the emission region. The spectral index of emitted radiation or observed photons depends on the confinement 
time of the particles within acceleration region. Therefore, the observed variability in the blazar light curves can be attributed to 
the fluctuations in the acceleration rate of electrons under leptonic scenario.

\subsection{Geometrical Effects}
Geometrical effects like varying Doppler factor due to changes in the viewing angle of the emission region in the jet are used to interpret 
the long term variability of the blazars \citep{Raiteri2017}. An inhomogeneous and curved jet can undergo dynamical orientations due to the 
magnetohydrodynamic instabilities or rotation of the twisted jet. This leads to the changes in the orientation of the jet emission regions 
and hence their Doppler factors. Therefore, emissions from these regions in the jet are enhanced or Doppler boosted when their 
orientation is aligned with the line of sight of the observer. These models account for the continuous time evolution of the broadband 
emissions from an inhomogeneous, curved jet. Shortening of the variability timescales during the flaring episodes of blazars is an 
important observational feature of the long term variability due to changing Doppler factor.

\section{Temporal Variability Characterization}\label{sec3}
Time structure of the flux variations plays and important role to investigate the physical processes involved in the blazar variability. 
Therefore, temporal characterization of the flaring episodes in the multi-wavelength light curves of blazars is very crucial in the 
variability studies. Following methods are widely used to measure the characteristic variability timescale ($\tau$) which is defined as 
the time required for the flux to change by a given (constant) factor.

\subsection{Structure Function}
The structure function (SF) approach is used to estimate the characteristic variability timescale in the light curves of blazars 
\citep{Simonetti1985}. For a given light curve described as a time series $F(t)$ with the time lag $\tau$, the first order SF is 
defined as 
\begin{equation}
	SF(\tau)~=~\frac{1}{N}\Sigma_{i=1}^{N}(F_i - F_{i+\tau})^2
\end{equation}	
where $F_i$ and $F_{i+\tau}$ are the discrete flux points at given times $t$ and $t+\tau$ respectively, $N$ is the number of 
data points in the light curve separated by time interval $\tau$. The SF values are estimated as a function of $\tau$ and its look 
for a given light curve depends on the noise level and time binning for producing the light curve, and minimum and maximum timescales 
of the variability \citep{Emmanoulopoulos2010}. For very small values of $\tau$ (less than the shortest timescale), the SF is nearly 
constant due to noise dominance. For large values of $\tau$ (but less than the shortest variability timescale), SF is proportional to 
$\tau^2$ and characterizes small scale linear behavior in the light curve. For $\tau$ close to the characteristic variability timescale, 
SF indicates the variability present in the light curve and the peak of SF corresponds to the variability timescale for the light curve. 
For the values of $\tau$ larger than the characteristic timescale, SF forms a plateau. If the flux points in the light curve are not 
evenly sampled, the SF is computed using interpolation algorithms. The value of SF represents a running variance of the process which 
can involve range of timescales contributing to the variability in a light curve. Therefore, this method is generally employed only for 
comparison of variability in the light curves since the estimated value of $\tau$ may not be a reliable characteristic variability 
timescale \citep{Emmanoulopoulos2010}.

\subsection{Correlation Function}
The correlation function (CF) is used to investigate the minimum variability timescale by correlating the 
multi-wavelength light curves. The well known methods for the estimation of correlation function are 
interpolated CF (ICF) and discrete CF (DCF). The ICF method calculates value of CF with no error bar using the 
interpolation between the observed flux points and may be unreliable for under sampled light curves \citep{Gaskell1987}. 
Whereas the DCF method allows estimation of CF using all pairs of the flux points in the light curve 
and without any interpolation or invention of artificial data \citep{Edelson1988}. For a given light curve, 
the auto correlation function (ACF) using DCF method is given by 
\begin{equation}
	ACF(\tau)~=~\frac{1}{N}\Sigma_{i,j}^{N}\frac{(F_i - \bar F)(F_j - \bar F)}{\sigma^2_F}
\end{equation}	
where $F_i$ and $F_j$ form the pair of discrete flux points at times $t_i$ and $t_j$ respectively such that 
the pairwise lag ($t_j$ - $t_i$) is inside the time lag $\tau$ and $N$ represents number of pairs. $\bar F$ and 
$\sigma_F$ are the mean and standard deviation of the light curve. Repeated peaks in the shape of ACF as a function of 
time lag indicate characteristic variability timescale and value of $\tau$ where ACF crosses zero first time 
(zero-crossing time) characterizes the maximum correlation. It means the typical variability timescale is quantified 
as the shortest time it takes the ACF value to fall to zero.

\subsection{Doubling/Halving Time}
The doubling or halving time is estimated to obtain the minimum variability timescale for the light curves in which 
rapid variations are clearly evident during the flaring episodes. The doubling or halving time is mathematically defined 
by different relations. For pairs of flux points $F_i$ and $F_j$ observed at consecutive time $t_i$ and $t_j$ respectively, 
the doubling time ($T_2$) is defined as \citep{Zhang1999}
\begin{equation}
	T_2^{ij}~=~\frac{F_i + F_j}{2}\left| \frac{t_j - t_i}{F_j - F_i} \right|
\end{equation}	
and the characteristic variability timescale $\tau = min(T_2^{ij})$ for all the pairs in a light curve. It is obvious from 
the above expression that $T_2^{ij}$ depends on time binning or sampling of the light curve and signal-to-noise ratio of the 
flux measurement. If the error on $T_2^{ij}$ (obtained through simple error propagation on $F_i$ and $F_j$) for a given pair 
is larger than 20$\%$, the value of $T_2^{ij}$ can be omitted for estimation of the shortest variability timescale \citep{Zhang1999}.
Sometimes, the time needed for the flux to increase or decrease by a factor of 2 is evaluated to characterize the observed 
variability timescale. This is referred to as characteristic doubling/halving time ($\tau_2$) depending on the increase or decrease 
in the flux and therefore can be defined as \citep{Brown2013}
\begin{equation}
	F_j~=~F_i~2^{(t_j-t_i)/\tau_2}
\end{equation}	
If $\tau_{2r}$ and $\tau_{2f}$ are the flux doubling/halving timescales during rise and fall of a flare respectively, then the 
flare superposed on stable or constant emission can be fitted by a function of the form 
\begin{equation}
	F(t)~=~F_c + \frac{F_0}{2^{(t_0 - t)/\tau_{2r}}+2^{(t - t_0)/\tau_{2f}}}
\end{equation}	
where $F_c$ is constant flux and $F_0$ corresponds to the highest flux level during flare at time $t_0$. During the fitting 
of the flare profile, $F_c$, $F_0$, $\tau_{2r}$ and  $\tau_{2f}$ are left as free parameters whereas $t_0$ is fixed to 
time corresponding to the peak flux in the light curve. The characteristic variability timescale of the flare is given by 
$\tau_2 = min(\tau_{2r},\tau_{2f})$.

\subsection{e-folding Time}
The e-folding time is defined as the characteristic time required for the flux to change by a factor $e^{\pm}$. 
In this method, it is assumed that the flux points in a light curve evolve exponentially during the rise or decay of the flare. 
For statistically significant pair of fluxes $F_i$ and $F_j$ (i $>$ j), the e-folding time ($T_e$) is defined as 
\citep{Burbidge1974,Calderone2011}
\begin{equation}
	T_e^{ij} = \left|\frac{t_i - t_j}{\rm ln F_i -\rm ln F_j}\right|
\end{equation}	
and the minimum e-folding variability timescale is estimated as $\tau_e = min(T_e^{ij})$. This method uses only two flux measurements 
in the light curve and no fitting is required for calculating the variability timescale. The resulting values of $T_e^{ij}$ are described 
by a normal distribution. The complete time profile of an exponentially rising and falling flare can be reproduced using the function 
\citep{Abdo2010b}
\begin{equation}
	F(t)~=~F_c + \frac{F_0}{e^{(t_0 - t)/\tau_{er}} + e^{(t - t_0)/\tau_{ef}}}	
\end{equation}
where $\tau_{er}$ and $\tau_{ef}$ are the e-folding rise and fall timescales respectively. Other parameters are same as defined in 
Equation (8). The characteristic e-folding variability timescale is obtained as $\tau_e = min(\tau_{er},\tau_{ef})$. The equivalent 
doubling timescale is given by 
\begin{equation}
	\tau_2 = \tau_e \times \rm ln 2
\end{equation}	
The epoch corresponding to the peak of the flare ($t_{peak}$) can be estimated from the maxima of $F(t)$ i.e.,
\begin{equation}
	\frac{d}{dt}F(t)~=~0
\end{equation}	
which gives
\begin{equation}
	t_{peak}~=~t_0 + \frac{\tau_{er}\tau_{ef}}{\tau_{er} + \tau_{ef}}\rm ln\left(\frac{\tau_{ef}}{\tau_{er}}\right)
\end{equation}	
and a good estimate of the total duration of a flaring episode is 
\begin{equation}
	t_{flare}~\approx~2(\tau_{er} + \tau_{ef})
\end{equation}	

\subsection{Linear Function}
If the flux-time dependence can be described by a linear relation, the timescale of flux variation is defined as \citep{Montagni2006}
\begin{equation}
	\tau~=~\frac{\bar F}{dF/dt}
\end{equation}
For constant value of $\tau$ in a given light curve, this corresponds to the e-folding timescale. For optical light curves with magnitudes 
$m(t)$, $\tau$ can be directly computed using the relation
\begin{equation}
	\tau~=~1.086\left(\left|\frac{dm}{dt}\right|\right)^{-1}
\end{equation}	
This is known as the classic Pogson's formula of the astrophysical photometry. A given light curve is divided into monotonic intervals to 
obtain the values of $\tau$. In each interval, the flux points are fitted with a straight line and the best fit slope value is assigned as 
the value of $\frac{dm}{dt}$. The bias introduced due to the subjective selection criterion for the intervals can be avoided using small 
number of intervals and long time intervals to contain sufficient number of data points.

\section{Variability Quantification}
The variability observed in a given light curve is a convolution of the intrinsic physical processes in the source and 
residual measurement errors introduced during the observations. Therefore, it is very important to quantify the level of 
variability present in the light curve to constrain the statistical and physical properties intrinsic to a source with 
variable emissions like blazars. Following parameters are defined to characterize the variability in the multi-wavelength 
light curves of the blazars.

\subsection{Reduced-$\chi^2$}
A $\chi^2$-test of a null hypothesis for constant emission model of the light curves is performed to characterize the variability. 
The reduced- $\chi^2$ ($\chi_r^2$) of the null hypothesis is defined as 
\begin{equation}
	\chi_r^2~=~\frac{1}{N}\Sigma_{i=1}^N\left(\frac{F_i - \bar F}{\sigma_i}\right)^2
\end{equation}	
where $\sigma_i$ is the error in the flux measurement $F_i$. If the probability of $\chi_r^2$ corresponding to $N-1$ degrees of freedom 
is $\le~0.01\%$, the light curve is considered as variable. The best fit constant flux in the absence of variability (probability 
of $\chi_r^2~ \ge~ 99.99\%$) is given by 
\begin{equation}
	F_c~=~\frac{\Sigma_{i=1}^N\frac{F_i}{\sigma_i^2}}{\Sigma_{i=1}^N \sigma_i^{-2}}
\end{equation}

\subsection{C-Parameter}
C-parameter is commonly used to test for the presence of variability in a light curve. It is defined as \citep{Jang1997}
\begin{equation}
	C~=~\frac{\sigma_{LC}}{\sigma}
\end{equation}	
where $\sigma_{LC}$ is the standard deviation of the flux points in the light curve and $\sigma$ is the average of 
nominal errors ($\sigma_i$) associated with each flux point. A light curve is characterized as variable at 99$\%$ 
confidence level if the value of C-parameter is above the critical value of 2.576. The variability quantification on the 
basis of C-parameter is questioned because the C-statistics does not follow a Gaussian distribution and the critical value 
of 2.576 is very conservative \citep{Diego2010}.

\subsection{F-Parameter}
The statistics based on F-parameter is used for assigning the significance to the observed variability. 
For a given light curve, the F-parameter is expressed as \citep{Howell1988},
\begin{equation}
	F~=~\frac{\sigma_{LC}^2}{\sigma^2}
\end{equation}	
where $\sigma_{LC}^2$ is the variance of the light curve, and $\sigma^2$ is the mean square error. The square 
value of the C-parameter can be used to estimate the unbiased value of F-parameter. If F exceeds a critical value 
of 6.636 for a chosen significance level corresponding to (N-1) degrees of freedom, the null hypothesis 
(no variability) is rejected \citep{Diego2010}. Because F-parameter is defined as the ratio of two variances, it 
describes a normally distributed variable. Therefore, the F-parameter can be considered better than C-parameter for 
reliability of the variability present in a light curve.

\subsection{ANOVA Test}
The analysis of variance (ANOVA) test is performed to investigate the presence of IDV or micro-variability in the 
light curves of blazars \citep{Diego1998,Diego2010}. The search for variability using ANOVA test does not depend on the measurement error 
unlike the F-parameter. In the ANOVA test, a light curve is divided into various groups with appropriate number of data points according 
to its length and time sequence. If last group has less data points, it is merged with the previous group. The total deviations of the 
light curve can be separated into total variations between and within groups. Variance of the means of each group and mean variance for 
the dispersion within the groups are computed. The ratio of these two variances multiplied by the number of data points in each group 
follows F-statistics. Therefore, the critical value for ANOVA test ($F^{\beta}_{\nu_1,\nu_2}$) can be derived from the F-distribution on 
the basis of significance level ($\beta$), degrees of freedom for the groups ($\nu_1 = k-1$), and degrees of freedom within the 
groups for errors or dispersion ($\nu_2 = N-k$) where $N$ is the total number of data points in a light curve and $k$ is the 
number of groups. If the number derived from the ANOVA test exceeds the critical value ($F^{\beta}_{\nu_1,\nu_2}$) for a given 
significance level, the null hypothesis (no variability) is rejected. The null hypothesis for ANOVA test is that the means of 
different groups are equal. If the test yields a probability less than the adopted significance level, alternate hypothesis 
that at least one group mean is different will be accepted. The alternate hypothesis indicates presence of variability in the light 
curve of blazar. A detailed mathematical treatment of the ANOVA test can be found in \citep{Diego2010}. Tests for means are found to be 
more powerful than tests for variances. Therefore, the ANOVA test is more robust and powerful than the F-parameter method for detecting 
the variability at minutes timescale in a light curve. However, a robust use of ANOVA test requires sufficiently large number of data 
points in a light curve.

\subsection{Enhanced F-Parameter}
The enhanced F-parameter is employed to detect the presence of variability in the optical and infrared light curves having 
a substantial brightness difference between blazar and comparison stars \citep{Diego2014,Diego2015}. A large brightness mismatch or 
significant variability in a comparison star may lead to the underestimation of the blazar variability using the standard F-parameter. 
To overcome this problem, the variance of differential light curve of comparison stars is enhanced or scaled up by an appropriate factor. 
The enhancement factor ($\kappa$) is defined as the ratio of mean square error in the differential light curve for blazar and one comparison star 
to the mean square error in the differential light curve of two comparison stars. Using this the enhanced F-parameter is estimated as 
\begin{equation}
	F^{en}~=~\frac{\sigma^2(blazar - star1)}{\kappa~\sigma^2(star1-star2)} 
\end{equation}
where $\sigma^2(blazar - star1)$ and $\sigma^2(star1-star2)$ are the variances of the differential light curves for 
blazar-comparison star 1 and comparison stars 1 \& 2 respectively. The value of $F^{en}$ is compared with the critical F-value to 
decide the presence of variability in the blazar light curve. The basic idea for enhanced F-parameter is to transform the 
differential light curve of comparison stars to have the same measurement noise, as if their brightness exactly matched the mean 
brightness of the blazar. The use of multiple standard stars reduces the probability of false variability as compare to a single 
comparison star.   

\subsection{Variability Index}
The variability index (V) is estimated to quantify the level of variability through the measurement of peak-to-trough variation 
of the flux in a given light curve. It is defined as \citep{Aller1992}
\begin{equation}
	V~=~\frac{(F_{max} - F_{min}) - (\sigma_{max} + \sigma_{min})}{(F_{max} + F_{min})-(\sigma_{max} - \sigma_{min})}
\end{equation}
where $F_{max}$ and $F_{min}$ are the highest and lowest flux values respectively, and $\sigma_{max}$ and $\sigma_{min}$ 
are the associated measurement uncertainties. The use of flux errors in the definition of variability index underestimates 
the value of $V$, but greatly reduces the effect of fluctuations in the measurement on the intrinsic variability of the source.
The parameter $V$ indicates a significant variability only when the intrinsic variability of the source dominates the 
measurement uncertainties. For light curves obtained from the observations of a source with low signal-to-noise ratios, the 
variability index parameter can be negative or indicates very low intrinsic variability. A similar parameter, generally known as 
relative variability amplitude (RVA) is also defined to characterize the peak-to-trough variability in the light curves. The 
\emph{RVA} is estimated using the relation \citep{Kovalev2005}
\begin{equation}
	RVA~=~\frac{F_{max} - F_{min}}{F_{max} + F_{min}}
\end{equation}	
and the uncertainty associated with \emph{RVA} is given by \citep{Singh2018}
\begin{equation}
	\Delta RVA=\frac{2}{(F_{max}+F_{min})^2}\sqrt{(F_{max}~\sigma_{min})^2 + (F_{min}~\sigma_{max})^2}
\end{equation}	
For the statistically significant variability in a light curve, \emph{RVA} $\ge$ 3$\times \Delta$\emph{RVA}.

\subsection{Variability Amplitude}
The variability amplitude parameter ($A_{mp}$) is defined to quantify the actual variation in a light curve after 
correcting for the observational errors. It is calculated using the following formula \citep{Heidt1996}
\begin{equation}
	A_{mp}~=~\sqrt{(F_{max} - F_{min})^2 - 2\sigma^2}
\end{equation}	
and the percentage variation in the variability amplitude is given by \citep{Romero1999}
\begin{equation}
	A_{mp}(\%)~=~\frac{100}{\bar F}\times \sqrt{(F_{max} - F_{min})^2 - 2\sigma^2}
\end{equation}
The error in $A_{mp}$ is estimated as \citep{Singh2018}
\begin{eqnarray}
       \Delta A_{mp}(\%)~=~100\times \left(\frac{F_{max}-F_{min}}{\bar F \times A_{mp}}\right) \times \\
	                   \sqrt{\left(\frac{\sigma_{max}}{\bar F}\right)^2 
	                   + \left(\frac{\sigma_{min}}{\bar F}\right)^2 + \left(\frac{\sigma_{\bar F}}{F_{max}-F_{min}}\right)^2 A_{mp}^4}
\end{eqnarray}
where $\sigma_{\bar F}$ is the error in mean flux. Like variability index parameters, $A_{mp}$ also characterizes the peak-to-peak variation 
in the light curve. For optical light curves, $\sigma$ should be scaled up by a factor of 1.54 to take account the under estimation of the 
photometric errors \citep{Goyal2013}.

\subsection{Modulation Index}
The modulation index parameter (m) provides a measure of the observed variability amplitude without considering the 
error in the individual flux measurement. It is defined as the ratio of the standard deviation ($\sigma_F$) of the flux 
points to the average or mean ($\bar F$) of the fluxes in a given light curve and can be expressed as \citep{Quirrenbach2000}
\begin{equation}
	m~=~100 \times \frac{\sigma_F}{\bar F}~~\%
\end{equation}	
where
\begin{equation}
	\sigma_F~=~\sqrt{\frac{1}{N}\Sigma_{i=1}^N\left(F_i - \bar F\right)^2}
\end{equation}	
and 
\begin{equation}
	\bar F~=~\frac{1}{N}\Sigma_{i=1}^N F_i
\end{equation}
The modulation index (m) is always greater than zero and therefore can be considered as superior to the variability index.
However, it is a convolution of the variability intrinsic to the source, measurement uncertainties and effects of 
finite sampling. $m$ is also referred to as raw modulation index. The large value of $m$ indicates either strong variability 
during high emission states of the source or low activity state with large measurement errors. Therefore, exact interpretation 
of modulation index depends on the proper estimation of the measurement errors in the individual flux points. In order to estimate 
the true amplitude of the variability in a light curve, intrinsic modulation index ($\bar m$) has been defined \citep{Richards2011}. 
The basic principle of intrinsic modulation index is same as the standard modulation index except that $\bar m$ is estimated for 
zero measurement uncertainties and perfectly uniform sampling of the light curve or infinite number of samples. 
It is expressed as \citep{Richards2011}
\begin{equation}
	\bar m~=~100 \times \frac{\sigma_{F_0}}{\bar F_0}~~\%
\end{equation}	
where $\sigma_{F_0}$ and $\bar F_0$ are the standard deviation and mean of the distribution of the fluxes with zero 
observational error in the light curve. The values of $\sigma_{F_0}$ and $\bar F_0$ can be estimated using a likelihood 
method assuming that the observed flux points in the light curve follow a normal distribution with Gaussian errors. A 
detailed description of the likelihood approach can be found in \citep{Richards2011}. This method provides an estimate 
for the error associated with the intrinsic modulation index parameter, which is very useful for the comparison of different types 
of sources. The modulation indices of the multi-wavelength light curves of a source can be compared using a variability amplitude 
parameter, defined as \citep{Quirrenbach2000}
\begin{equation}
	Y~=~3\sqrt{m^2-m_0^2}~~\%
\end{equation}	
where $m_0$ is the modulation index of a non-variable or steady source observed with the same experiment. The factor 3 is arbitrarily 
chosen to make $Y$ compatible with the peak-to-peak variability amplitude. By definition, $Y$ should be zero for non-variable blazars.

\subsection{Fractional Variability Amplitude}
The uncertainties on the individual flux measurements in a light curve contribute an additional variance in the estimation of 
source variability. Therefore, in order to quantify the intrinsic variability of the source, the effect of measurement noise 
(Poisson noise in the photon counting measurements) should be corrected. An \emph{excess variance} is defined as an estimator of 
the intrinsic source variance which is the variance of the light curve after subtracting the contribution due to measurement 
errors \citep{Nandra1997,Edelson2002}. The fractional variability amplitude ($F_{var}$) parameter is defined as the square 
root of the normalized excess variance (ratio of intrinsic excess variance to the square of mean flux) to measure the 
level of variability intrinsic to the source. The $F_{var}$ is expressed as \citep{Vaughan2003} 
\begin{equation}
	F_{var}~=~\sqrt{\frac{S^2 - \sigma^2_{err}}{\bar F^2}}
\end{equation}
where
\begin{equation}
	S^2~=~\frac{1}{N-1}\Sigma_{i=1}^N (F_i - \bar F)^2
\end{equation}
is the total variance of the flux points and 
\begin{equation}
	\sigma^2_{err}~=~\frac{1}{N}\Sigma_{i=1}^N \sigma^2_i
\end{equation}
is the associated mean squared error. The parameter $F_{var}$ is described by a linear statistic and can give the 
percentage of the variability amplitude. 
The uncertainty on  $F_{var}$ estimated using standard error propagation formula is valid only for the uncorrelated 
Gaussian processes whereas the blazar light curves are strongly correlated and non-Gaussian \citep{Vaughan2003}. Therefore,
computational approaches based on Monte Carlo simulations of the red noise light curves are developed to estimate the effect 
of measurement errors on the values of $F_{var}$. The formal error in $F_{var}$ derived from the Poisson noise induced 
uncertainty on the excess variance is given by \citep{Vaughan2003,Poutanen2008}
\begin{equation}
	\Delta F_{var}~=~\sqrt{F_{var}^2 + \sqrt{\frac{2}{N}\left(\frac{\sigma^2_{err}}{\bar F^2}\right)^2 + 
			  \frac{\sigma^2_{err}}{N}\left(\frac{2 F_{var}}{\bar F}\right)^2}} -F_{var}
\end{equation}	
This equation is valid for the fluxes with errors described by both Gaussian and Poisson distributions. Thus, characterization of 
variability amplitude for a light curve using ($F_{var} \pm \Delta F_{var}$) takes into account the fluctuation  or uncertainty in 
the flux measurement and does not consider the intrinsic scatter in the fluxes inherent in any red noise process. A negative 
intrinsic excess variance or imaginary value of $F_{var}$ indicates no variability in the light curve or over estimation of the 
measurement errors. It is also important to note here that the estimated value of $F_{var}$ strongly depends on the temporal bin 
and sampling of the light curve. For a light curve with small temporal bins and dense sampling, the value of $F_{var}$ can be 
higher than that for the larger time bins due to smoothing out the short term variability. Therefore, the short-timescale variability 
is quantified using a similar parameter, known as point-to-point fractional variability amplitude ($F_{pp}$). 
It is defined as \citep{Edelson2002}
\begin{equation}
	F_{pp}~=~\frac{1}{\bar F}~\sqrt{\frac{1}{2(N-1)}\Sigma_{i=1}^{N-1}(F_{i+1}-F_i)^2-\sigma^2}
\end{equation}	
This parameter measures the variability between the adjacent flux points in a light curve. For white noise processes, $F_{var}$ and 
$F_{pp}$ will have same values, whereas $F_{var}$ will be higher than $F_{pp}$ for red noise due to large variations on long timescales. 
 
\subsection{Power Spectral Density}
The variability randomly observed in blazar light curves can be considered as a noise produced by a stochastic process and 
not due to a deterministic process. This means broadband emissions from blazars are the output of a noise process intrinsic to the 
source. The fluctuation power spectral density (PSD) is derived to characterize the variability of blazars and noise processes 
in general. The PSD represents the amount of power in the variability (defined as the average of squared amplitude) of emission 
as a function of temporal frequency or variability timescale. For a light curve described by real flux points $F(t)$, the PSD is given 
by \citep{Press1986} 
\begin{equation}
	P(f)~=~2\left|\hat F(f)\right|^2~~~~~; 0 \le f \le \infty
\end{equation}
where $P(f)$ is the power at the signal temporal frequency $f$ (inverse of the variability timescale) and $\hat F(f)$ is the Fourier 
transform of the light curve $F(t)$. For a light curve comprising a time series of fluxes $F_i$ measured at discrete times $t_i$, the 
discrete Fourier transform at each sampled frequency $f$ is given by 
\begin{equation}
	\hat F(f)~=~\Sigma_{i=1}^N F_i e^{2\pi i f t_i}
\end{equation}	
The mean flux $\bar F$ is subtracted from the light curve $F_i(t_i)$ to eliminate the zero-frequency power before estimating the 
Fourier transform $\hat F(f)$. The PSD for a light curve $P(f)$ can be computed using Fourier transform multiplied by its complex 
conjugate. However, computation of PSD in practice is very challenging due to irregular and odd sampling combined with the finite 
nature of the real data points in the light curves. Several popular algorithms for computation of the PSD can be found in the 
literature \citep{Scargle1982,Uttley2002,Marshall2015}. For blazars, the PSD is described by a power law of the form 
\begin{equation}
	P(f)~\propto~f^{-\alpha}~~~~~; 1 \le \alpha \le 3
\end{equation}	
where $\alpha$ is the PDS spectral slope. This indicates that variability amplitude will be higher on longer timescales 
(lower values of $f$) for $\alpha > 0$ in a given energy band. Red noise processes are described by uncorrelated fluctuations 
where PSD decreases with increasing frequency ($\alpha \ge$ 1). The true value of the variance of a light curve is given by 
integral of the PDS spectrum. For a light curve (a discrete time series), the observed variance is given by \citep{Vaughan2003}
\begin{equation}
	\sigma_{LC}^2~=~\Sigma_{i=1}^{N/2} \frac{P(f_i)}{N\Delta t}
\end{equation}	
where $f_i= i/N \Delta t$ is the frequency sampled by discrete Fourier transform and $\Delta t$ is sampling period. The minimum 
and maximum sampling frequencies are $1/N\Delta t$ and $1/2\Delta t$ (Nyquist frequency) respectively.

\section{Discussion and Conclusion}
The characteristic timescales from temporal variability and variability parameters obtained from the multi-wavelength 
light curves of blazars can be used to probe the intrinsic emission  properties like radiative processes, size and location 
of the emission region in the jet and mass of the SMBH at the centre of the host galaxy. The long term multi-wavelength 
variability can not be probed through observations because the blazars have been observed for nearly two decades only. 
The origin of long term variability in the emission is most probably linked to the variations in the accretion rates 
onto the SMBH. The short term variability is generally attributed to the propagation of shocks down the jet whereas 
the intra-day/night variability is caused due to the turbulence in the plasma outflow in the inner-most region of the jet. 
The rapid variability in the broadband radiation suggests relativistic expansion of the jet and the observed emission is 
dominated by specific regions in the large scale jets. The variability at minute timescales suggests a very compact emission 
and the causality arguments can be used to place an upper limit on the minimum size of the emission region in the jet given by 
\begin{equation}
	R~\le~\frac{\tau~c~\delta}{1+z}
\end{equation}	
If the variability originates in the jet formation region, its size should be of the order of the Schwarzschild radius of 
the SMBH and the observed minimum variability timescale can be given by the event-horizon light crossing time. In this case, 
the observed luminosity and the minimum variability timescale are related by Elliot-Shapiro (E-S) relation \citep{Elliot1974} 
\begin{equation}
	\rm log~L \le \rm log~\tau + 43.1
\end{equation}	
where $L$ is luminosity in units of erg~$s^{-1}$ and variability timescale $\tau$ in the units of second. The variability timescale 
also constrains the gravitational radius of the SMBH. This results in a crude estimation of the mass of SMBH which is an important 
parameter to probe the blazar phenomena. Assuming that the observed minimum variability time-scale is manifested by the orbital period 
of the innermost stable orbit around the maximally rotating SMBH, its mass is given by \citep{Xie2002}
\begin{equation}
	M_{BH}~=~1.62 \times 10^4 \frac{\tau~\delta}{1+z}~M_\odot
\end{equation}	
where the minimum variability timescale $\tau$ is in the units of second and $M_\odot$ is the solar mass. Thus, the minimum 
variability timescale can provide important information about the size of emission region and the mass of SMBH at the center 
of blazar host galaxy. If the variable non-thermal emission originates from the vicinity of the SMBH, a correlation between 
the variability timescale and mass of black hole is expected. In the absence of any correlation, the variable emission in the jet 
can be produced at large distance from the SMBH. For a relativistic jet with conical shape, the location of variable emission 
zone ($r$) from the central region is given by 
\begin{equation}
	r~=~\frac{R}{\rm tan~ \theta}
\end{equation}
The time-lag between the peaks of a multi-wavelength flare indicates the relative location of the emission zones and associated physical 
mechanisms including particle acceleration and energy dissipation in different energy bands. The geometry of jet magnetic field and 
material content of the relativistic jet its dynamics can also be probed using multi-wavelength variability of the blazars. 
The observed time-lag between the optical and $\gamma$-ray flares ($\Delta \tau$) is related to location of $\gamma$-ray dissipation 
region in the jet as $r~\sim~ c~\delta^2~\Delta \tau$. 
\par
The variability amplitude parameters are generally found to be anticorrelated with the luminosity and correlated with the mass 
of SMBH. $F_{var}$ and $F_{pp}$ estimated for the simultaneous multi-wavelength light curves describe the energy dependence of the 
observed variability at long and short timescales respectively. If $F_{var}$ vs energy plot for multi-wavelength observation of a 
blazar during flaring episode shows a double hump structure with peaks at X-ray and $\gamma$-ray energies, it indicates a correlation 
between synchrotron and inverse Compton emissions and therefore supports the leptonic origin of the high energy emission from blazars. 
The ratio ($\frac{F_{var}}{F_{pp}}$) as a function of energy or frequency measures the dependence of PSD slope ($\alpha$) on energy. 
The large values of the ratio indicate steeper PSD spectrum. If the ratio does not depend on energy, $\alpha$ is also energy independent. 
Therefore, PSD is very useful in identifying the characteristic variability timescale and for the comparison of the LTV and STV in different 
energy bands. An empirical relation between the $F_{var}$ for X-ray light curve and $M_{BH}$ is given by \citep{Lu2001}
\begin{equation}
	M_{BH}~\approx~\frac{10^5}{F_{var}^2}~M_\odot
\end{equation}
Therefore, the observed minimum variability timescale from multi-wavelength observations and $F_{var}$ estimated for 
the X-ray light curve can be used to impose a stringent constrain on the mass of SMBH in the blazar host galaxy. 
The physical evolution of the flares in the blazar light curves is probed using a parameter defined as \citep{Abdo2010b}
\begin{equation}
	\zeta~=~\frac{\tau_{ef} - \tau_{er}}{\tau_{ef} + \tau_{er}}
\end{equation}
with the associated error given by \citep{Singh2020} 
\begin{equation}
	\Delta \zeta~=~\frac{2}{(\tau_{ef} + \tau_{er})^2}\sqrt{(\tau_{ef}\Delta \tau_{er})^2+(\tau_{er}\Delta \tau_{ef})^2}
\end{equation}	
where $\Delta \tau_{er}$ and $\Delta \tau_{ef}$ are the uncertainties in $\tau_{er}$ and $\tau_{ef}$ respectively as defined in 
Equation (10).
The value of $\zeta$ lies between -1 and +1 indicating completely right and left asymmetric flares respectively. $\zeta = 0$ 
characterizes a symmetric flare with similar rise and decay timescales. The symmetric flares are related to the crossing time 
of photons or radiating particles through the dissipation region and may be a superposition of many short duration flaring 
episodes. The marked asymmetric flares with extreme value of $\zeta$ (close to $\pm$1) correspond to the fast injection of 
accelerated particles and slow radiative cooling or escape from the emission region.
\par
In summary, a detail study of the correlation between the observed features of the variability and certain properties of 
the emission region in the jet is very important to probe the physics of variability in blazars. The evolution 
of broadband SEDs of blazars on short timescales during flaring episodes at X-ray and $\gamma$-ray energies helps 
in understanding the particle acceleration in the jet. The physical mechanisms causing the unprecedented variability in 
the broadband emission from blazar jet are being investigated. However, some of the proposed physical processes intrinsic 
to the jet for the variable emissions are relativistic turbulence in the jet, cooling of relativistic particles produced 
by various acceleration and injection processes and intervention of shock waves responsible for acceleration. The external 
mechanisms like interactions of the jet with red giant stars through direct collisions and geometrical effects like changing 
viewing angle and Doppler factor are also invoked to understand the variability of blazars. The observed multi-wavelength 
emissions from blazars varying at long timescales can be interpreted as the modulation of changes in the orientation of 
different emission regions. The characterization of variability in the blazar light curves discussed here can play an important 
role in exploring the physics of blazars and other similar astrophysical sources in the vibrant era of multi-messenger astronomy.
\section*{Acknowledgments}
We thank the anonymous reviewer for his/her valuable suggestions to improve the manuscript.
\bibliography{MSR2}%
\end{document}